\documentclass{article}

\usepackage{PRIMEarxiv}

\usepackage[utf8]{inputenc} 
\usepackage[T1]{fontenc}    
\usepackage{hyperref}       
\usepackage{url}            
\usepackage{booktabs}       
\usepackage{amsfonts}       
\usepackage{nicefrac}       
\usepackage{microtype}      
\usepackage{lipsum}
\usepackage{fancyhdr}       
\usepackage{graphicx}       
\graphicspath{{figures/}}     

\usepackage{placeins}

\let\Oldsection\section
\renewcommand{\section}{\FloatBarrier\Oldsection}

\title{Deep Learning Architectures for FSCV, a Comparison
}

\author{
  Thomas Twomey$^1$, Leonardo Barbosa$^1$, Terry Lohrenz$^1$, P. Read Montague$^{1,2}$ \\
  $^1$Fralin Biomedical Research Institute, Virginia Tech, Roanoke, VA \\
  $^2$Department of Physics, Virginia Tech, Blacksburg, VA
}

\begin{document}
\maketitle

\begin{abstract}
We examined multiple deep neural network (DNN) architectures for suitability in predicting neurotransmitter concentrations from labeled in vitro fast scan cyclic voltammetry (FSCV) data collected on carbon fiber electrodes. Suitability is determined by the predictive performance in  the "out-of-probe" case, the response to artificially induced electrical noise, and the ability to predict when the model will be errant for a given probe. This work extends prior comparisons of time series classification models by focusing on this specific task. It extends previous applications of machine learning to FSCV task by using a much larger data set and by incorporating recent advancements in deep neural networks. The InceptionTime architecture, a deep convolutional neural network, has the best absolute predictive performance of the models tested but was more susceptible to noise. A naive multilayer perceptron architecture had the second lowest prediction error and was less affected by the artificial noise, suggesting that convolutions may not be as important for this task as one might suspect. 
\end{abstract}


\section{Introduction}
\begin{figure}[!htb]
  \centering
  \includegraphics[width=\textwidth]{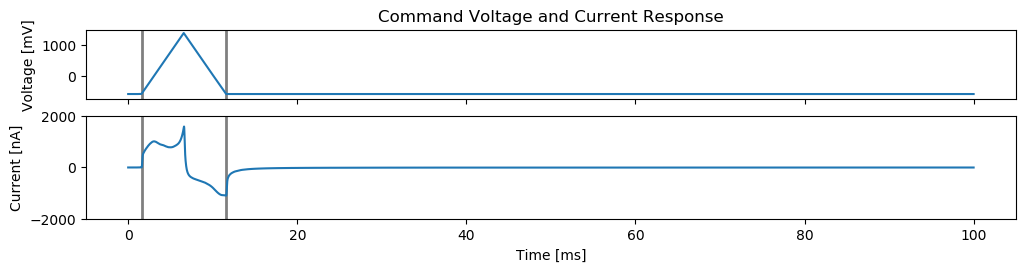}
  \includegraphics[width=\textwidth]{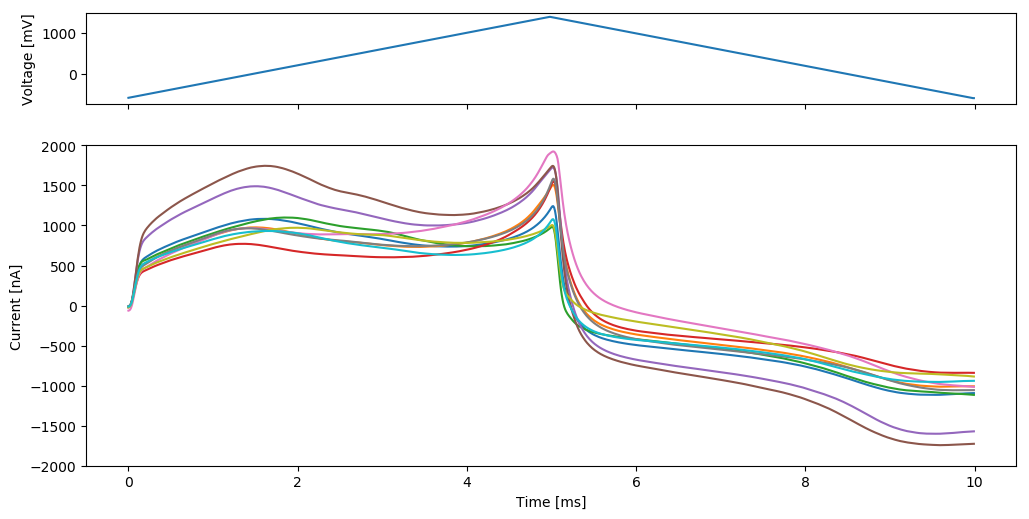}
  \caption{The command voltage and example current responses used in FSCV. The top plot shows a single 100 ms sweep with the active 10 ms section of the command and corresponding current denoted with vertical lines. The bottom plot of the 10 ms active section used by the models. The various color lines are the mean sweep for a probe averaged over all of its sweeps. }
  \label{fig:voltage_current}
\end{figure}

The function of the neurotransmitters Dopamine (DA), Serotonin (5HT), Norepinephrine (NE) in brains, and the role of their high frequency (> ~1 Hz) fluctuations are not well understood. Fast scan cyclic voltammetry (FSCV) using carbon fiber microelectrodes (probe) is the primary method used in rodents\cite{Phillips2003}\cite{Howe2013}, non-human primates\cite{Awake_Monkeys_2014}\cite{Primate2016}\cite{Monkey2015}\cite{Long_Term_Primates2017}, and, more recently, in humans \cite{kishida_2011}\cite{kishida_2015}\cite{Moran2018ER}\cite{BANG2020} to make sub-second measurements of these neurotransmitters. In FSCV a voltage waveform is applied to the probe and the invoked current is measured and used to make neurotransmitter concentration estimates. Example voltage and current waveforms are shown in Figure \ref{fig:voltage_current}. In this paper we consider the case of FSCV in awake humans during brain surgery. This application requires estimates of neurotransmitter concentrations to be generated from in vivo currents using models trained in vitro data sets. Inherit differences between individual probes and the instability \cite{kishida_2015}\cite{Loewinger2022} of their induced current responses motivates the use of “out-of-probe” test sets to validate these models. The average in vitro current response from a sample of probes in shown in Figure \ref{fig:voltage_current} and demonstrates these differences.

We viewed predicting concentrations from current traces as an amalgamation of a time series regression and an image regression task. The task, like prior work in the time series domain, has data that is sequentially gathered and thus exhibits high auto correlation. Additionally, some structures appear to move within the window of data collection. It differs from most time series work and is more similar to image-based tasks in that the observation is coincident in time with the stimulus. It differs from both domains as it is a regression task and only has a single channel input. Much work has been done quantifying the performance of various architectures for time series classification \cite{fawazDissertation} and we aimed to adopt and compare the successful architectures to this specific regression problem. Two architectures of note from that paper are the InceptionTime Network \cite{fawazInceptionTime} and the Fully Convolutional Network \cite{Wang_FCN}. In the time since that publication, more work has been done on time series classification of biological signals, incorporating advancements with transformer networks \cite{eeg_transformer}, and convolutional neural network derivatives of transformers \cite{SSVEP2022}. Prior attempts to use machine learning techniques have found success with principle component regression \cite{Loewinger2022}\cite{hitchhikers}, penalized regression models \cite{kishida_2015}\cite{Moran2018ER}, and DNNs with a different prepossessing approach\cite{dong2022}. 

To make this endeavor computationally feasible we constrained all models to have the same number of parameters and training procedure as the InceptionTime architecture \cite{fawazInceptionTime}. The InceptionTime architecture was chosen as the baseline as it had the best performance of the implemented models in \cite{fawazDissertation}. When possible values for hyperparameters were taken directly from the paper referenced. If it was not possible to use the same hyperparameter values, values were selected to match the parameter constraint. 

The various model architectures are evaluated to compare their performance regarding both relative correctness/linearity, and absolute correctness. Absolute predictive correctness is clearly important and allows direct comparison of true values across individuals or in the weaker case direct comparison of relative values across individuals. Relative predictive accuracy, a weaker requirement, is still useful. It allows for directional and proportional comparison across individuals. Further the confusion between DA and NE, where an increase in one analyte would cause an increase in prediction of the other analyte, is of extra note as previous work has had trouble with this problem \cite{Heien2003}.

We also consider how the models respond to electrical noise. In ideal conditions, data collection can be performed in a Faraday cage resulting in a low noise environment. In the human brain surgery case, medically required electronics induce small perturbations in the current. The largest of these perturbations occur at 60 hertz (The power grid frequency). This periodic noise has been observed to be either sinusoidal in nature or sharp with relatively large spikes. As deep neural networks have been shown to be effected by noise or perturbations in unintuitive ways \cite{API2017}\cite{CommonCorruptions20191}, the impact of this noise is of greater concern.

\section{Methods}

\subsection{Data sets}
Data was collected on 76 carbon-fiber probes manufactured as described in \cite{kishida_2015}. These probes were designed to be implanted in humans and run during deep brain stimulation implantation surgery. The four analytes considered in this paper are Dopamine (DA), Serotonin (5HT), Norepinephrine (NE) and pH. Typically, each analyte had its own data set collected on each probe. The data sets for the neuromodulators (DA, 5-HT, and NE) were collected with 30 concentrations of the neuromodulator of interest ranging from 0 to 2500 nanomolar (nM) with a pH of around 7.4, while the other neuromodulators had concentration of 0 nM. In addition, these data sets contained 5 of the mixture solutions in which all three analytes had a value of 0, 840, or 1680 nM. The pH data set was collected with 11 pH values in the range 7.0 to 7.8 and with the concentration of the three neuromodulators set to 0 nM as well as 5 mixture solutions as described above. The order of data collection was randomized with concentration in the data set and in the order of collection of the data sets.

This data collection approach was selected for practical and logistical reasons concerning the collection of the data. It results in suboptimal data sets with very skewed label distributions. For a given neuromodulator the label is 0 nM for approximately 70 percent of the samples and when its concentration is non-zero the other neuromodulators are zero for approximately 70 percent of the samples. 

The command voltage was applied as a triangle waveform with 90 milliseconds of a baseline potential of -0.6 V and a slope of 400 V/s for 5 milliseconds followed by a slope of -400 V/s for 5 milliseconds. The electrical current data was collected at 100,000 Hz such that 1000 samples aligned with the “active” part of the command voltage. These 1000 samples are the current trace or sweep and serve as the input to the models as shown in Figure \ref{fig:voltage_current}. Data was collected in a wet lab environment with care taken to minimize electrical noise from lab equipment, but for the most part, not in a Faraday cage.

\subsection{Data Preprocessing}
For each probe a subset of sweeps contiguous in time was selected from each recording to minimize the amount of sampled noise and maximize sweep stability. All current traces had the numerical first derivative taken as prior work had found slight improvements over models trained on the raw current traces \cite{kishida_2015}\cite{BANG2020}\cite{Loewinger2022}. A shifted z-score normalization function was applied to the concentration vector labels. The intention of this normalization was to equally weight the error of the analytes despite the values falling in different ranges. The inverse of the shifted z-score function is applied to the output of the model to get the final prediction in real space.

\subsection{Model Architectures}
The following models were run using Tensorflow \cite{tensorflow} and Keras\cite{chollet2015keras} implementations. If possible the implementation from the source paper was used directly. The code for the models and the associated infrastructure can be found at https://github.com/teptwomey/Deep\_Learning\_Architectures\_for\_FSCV. All models have the same final layer to produce the final predictions. This final layer is a fully connected or dense layer with a softplus activation function.

\subsubsection{InceptionTime}
The InceptionTime network is an adaptation of the Inception image classification architectures adapted / tuned for the problem of multivariate time series regression \cite{fawazInceptionTime}. As implemented, the InceptionTime architecture has two ResNet blocks \cite{resnet2016} with three internal convolution blocks. These convolutional blocks consist of four parallel convolutional layers with 32 filters and kernels of size 1, 10, 20, and 40. The output of parallel layers is stacked, batch normalized, and fed to a RELU activation layer. The final layer of the convolutional block is bottleneck convolutional layer with a 32 filters of kernel size 1. Inside each ResNet block there is a bottleneck convolutional layer with 32 filters and a kernel size of 1. The output of the bottle neck is fed to the connected convolution blocks and the output of those blocks is added to the input of the ResNet Block and run through a RELU activation. After the sequential ResNet blocks there is a global average pooling layer and the standard dense final layer. The InceptionTime paper uses a ensemble of 5 models, but we only used a single model.

\subsubsection{Fully Convolutional Network (FCN)}
The fully convolutional network architecture has blocks containing a convolutional layer, a batch normalization layer, and a RELU activation layer. Following these blocks there is a global average pooling layer and a softmax layer. The final layer is a dense layer with four outputs and a soft-plus activation. Two versions of the FCN were trained. One model with the hyperparameters from the reference \cite{Wang_FCN} that has three blocks with filter sizes [128, 256, 128]. This model is referred to as FCN\_ref and has 265,220 parameters. The other model, referred to as FCN\_wide, has the same blocks but with additional filters at each level [162, 321, 162] such that the number of parameters was similar to InceptionTime with 419,899. 

\subsubsection{SSVEPformer}
The SSVEPformer \cite{SSVEP2022} architecture was designed to classify Steady-State Visual Evoked Potential signals in a brain-computer interface context. It was particularly tailored to the problem of inter-subject prediction in a similar vein to our inter-probe prediction goal. It is composed of an initial layer that takes the discrete Fourier transform of the input and concatenates the real and imaginary components. It then has a channel combination block with a 1D convolutional layer to produce two channels from each channel of input, a layer normalization layer, a GELU activation layer and a dropout layer with p = 0.5. That channel combination block is followed by two sub-encoders that each contain a CNN module and a channel MLP module. The CNN module has a layer normalization layer, a 1D convolutional layer, a layer normalization layer, a GELU activation layer, a dropout layer with p = 0.5, and a residual layer that adds the output of the dropout layer and the input to the CNN module. The channel MLP module is composed of a layer normalization layer, a dense layer, a GELU activation layer, a dropout layer with p = 0.5, and a residual layer that adds the output of the dropout layer and the input to the CNN module. The output of the two sequential sub-encoders is feed to a MLP head block. This MLP head block flattens its input, has a dropout layer, followed by a dense layer, a layer normalization layer, GELU activation, and another dropout layer. The output of the MLP head is fed to the standard final dense layer with softplus activation. As implemented the SSVEPformer has  428,588 parameters. 

\subsubsection{EEG-Transformer}
We implemented a derivative of the EEG-Transformer model \cite{eeg_transformer}, which itself is a modification of the Vision Transformer (ViT) \cite{vit2020} design. The EEG-Transformer was tuned for multi-channel time series classification and in particular EEG signals, we adapted it for single channel input and regression. The first block of the implemented transformer architecture uses a convolution layer to create the 1D equivalent of patches from the input time series. After the convolution a class token and position embedding are added to the patch. The patch and its embeddings are then given to three successive transformer blocks. Each transformer block has an attention sub-block, a residual connection, layer normalization layer, a MLP block and a residual connection. The attention sub-block has a layer normalization layer, a Multi-Head Attention layer, and a dropout layer. The MLP sub-block has a dense layer, a GELU activation layer, a dropout layer, a dense layer and a final dropout layer.
The output of the final transformer block goes through a final layer normalization, a mean reduction is preformed and the standard dense final layer with softplus activation is applied. As implemented the EEG-Transformer has 426,394 parameters.

\subsubsection{Multi-Layer Perceptron (MLP)}
A simple Multi-Level Perceptron (MLP) was created with three hidden layers of 312, 256, and 128 nodes respectively. Each hidden layer had GELU activation. The final layer had 4 outputs and a softplus activation. No dropout or regularization were implemented. The implemented MLP model has 425,540 parameters. Two versions of the MLP were implemented, the MLP\_wide to match the number of parameters from the InceptionTime model, and the MLP\_big model with an extra dense layer and 491,332 parameters. The larger MLP\_big model was run after the MLP\_wide model had been tested to provide context for the unexpected performance.

\subsection{Model Comparison}
A probe based ten-fold cross validation experiment was performed to quantify and compare each architecture. The 76 probes were divided into 10 folds, 6 folds with 8 probes and 4 folds with 7 probes. One fold was held out as the test fold and the other 9 folds were pooled to create a training and validation set. The training set consisted of roughly 90 percent of the unique probe-concentrations with the remainder being placed in the validation set. Once the model was trained it was used to predict the held-out test set. That process was repeated such that each fold was held out and predicted as a test set. In the end every probe had predictions run on its corresponding sweeps from a model that had not seen that probe in training or validation. The data sets were created once for each fold and were used to train and test all the model architectures. As the number of probes in each fold was not the same and there was some variation in the amount of data collected on each probe, the size of the training and validation sets was not equal across folds.

All models were implemented in Keras \cite{chollet2015keras} and trained using the ADAM optimizer \cite{adam2014}, mean square error as the loss function, a batch size of 64, and an initial learning rate of 0.001. The learning rate was halved following 5 consecutive epochs with no improvement in validation loss. Each model was trained for 35 epochs. Following each epoch, the model was evaluated on the validation set and following the completion of training, the epoch with the best validation loss was retained as the final version of the model to be used for prediction and testing. Each model was trained was using a single NVIDIA Tesla V100 16GB GPU.

\subsection{Response To Noise}
\begin{figure}[!htb]
    \centering
    \includegraphics[width=\textwidth]{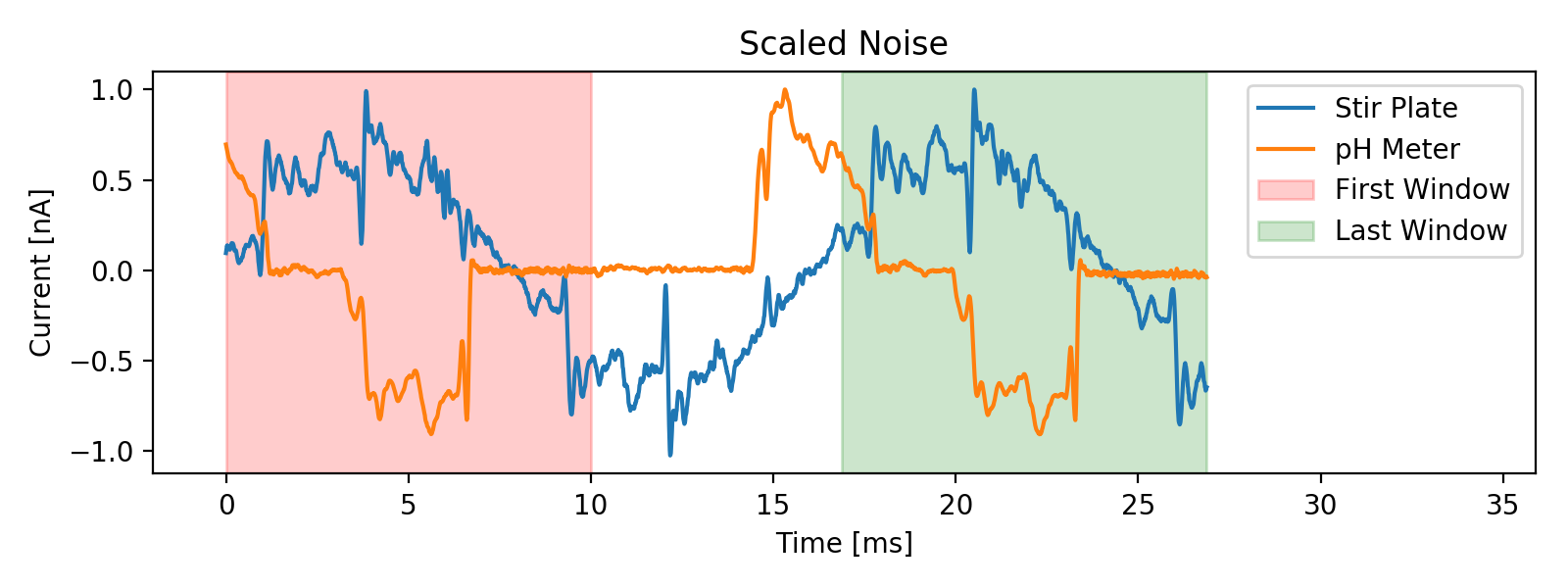}
    \caption{Isolated induced noise scaled to 1 nA peak. Shaded rectangles represent the first and last of the windows added to the sweeps. These window are generated via the moving window procedure.}
    \label{fig:example_noise}
\end{figure}

We conducted voltammetry recordings via our standard procedure in proximity to two commercial electronic devices, a magnetic stir plate and a bench top pH Meter. From the recordings were able to isolate the noise in the current recordings induced from these devices. These noise sources were selected because of their qualitative similarity to noise observed in in vivo environments. To validate that adding this isolated noise to existing current recordings was a valid procedure, we tested if the noise was additive. To test if the noise was additive we averaged full 100 ms sweeps directly before and directly after the noise source was turned on and subtracted the noise off average from the noise on average. The difference sweep had noise spikes of uniform magnitude across all current levels of the noise free sweep. For both noise sources, stir plate and pH meter, we isolated a 27-millisecond noise recording from the in vitro measurements. We than scaled both noise recordings such that the peak current was equal to six noise levels (0.25, 0.5, 0.75, 1, 1.5, 2 nA). The 1 nA scaled version of each noise source is shown in figure \ref{fig:example_noise}. 

The noise sources that we isolated are caused by the roughly 60 Hz power grid line frequency. This frequency is not exactly fixed at 60 Hz and continuously fluctuates in response to grid conditions. That fluctuation results in an apparent movement of the noise waveform across sequentially recorded current sweeps. To simulate this movement we sampled a window moving across the scaled noise recording. This moving window was equal in length to the input of the model (10 ms) and was shifted a single observation for each step. The window was moved 1667 steps to cover every position of the 60 Hz waveform. For every position, the corresponding sampled window was added to a reference clean 10 ms sweep. Figure \ref{fig:example_noise} shows the first and last positions of the moving window.

To conserve computational resource we constrained our noise investigation to the trained models and test probes from the first fold of our k-fold test. We then selected a single sweep from every unique neurotransmitter concentration from all the probes in the this fold. These selected sweeps act as our base sweeps. Each base sweep had noise added it to and predictions run. The noise was independently added from every combination of noise source, noise level, and from all positions of the moving window. Thus for every base sweep, ~20,000 (2 noise source x 6 noise levels x 1667 moving window locations) base-plus-noise sweeps had predictions run. We note that the base sweeps had some existing level of noise and with the addition of the artificial noise, some level of constructive and destructive interference could have occurred. This interaction may create an apparent noise form that may be qualitatively different from the artificial noise.

\section{Results}

\subsection{Predictive Performance}
\begin{figure}[!htb]
  \centering
  \includegraphics[width=\textwidth]{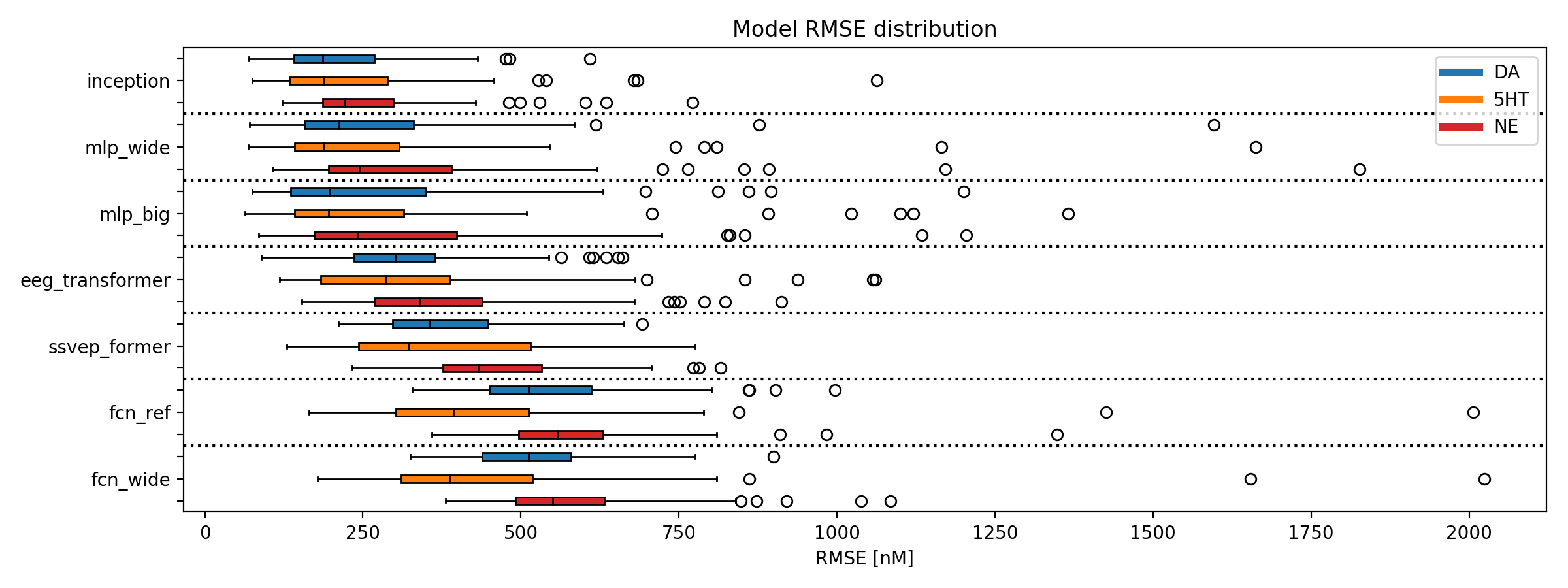}
  \includegraphics[width=\textwidth]{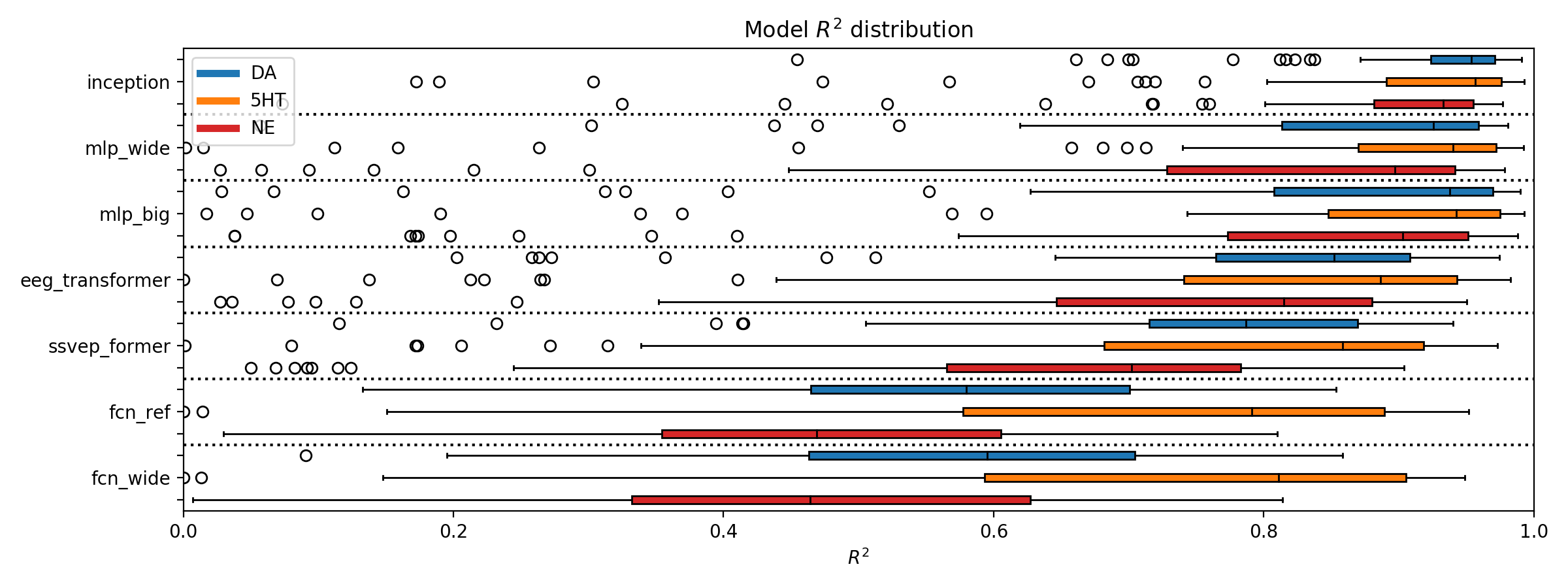}
  \caption{Root Mean Squared Error (RMSE) and Coefficient of Determination $R^2$ Box Plots for each model. Circles represent outliers falling more than 1.5 times the IQR away from the first or third quartile.}
  \label{fig:boxplots}
\end{figure}

The distribution of values for the root mean squared error (RMSE) and R-squared ($R^2$) of the linear fit of true and predicted concentration are shown in Figure \ref{fig:boxplots} for DA, 5HT, and NE (pH is omitted). The corresponding mean and standard deviation if these RMSE values are listed in Table \ref{tab:rmse_table}. As shown, of all the models considered, the InceptionTime architecture had the best median RMSE and $R^2$ values across all three neurotransmitters considered. Surprisingly, the MLP performed similarly in terms of the median and first quartile, but generally had a longer right tail. For the MLP and FCN architectures, where multiple versions with different numbers of parameters were trained, there seems to be little improvement with the larger models. Model performance on a given neurotransmitter is correlated with performance on the other two neurotransmitters and the 5HT has the highest performance across the models.
The ordinality of the model performance results are unchanged when considering the RMSE or $R^2$ metrics. The $R^2$ results do show a greater relative spread with more probes classified as outliers.

\begin{table}[!htb]
 \caption{Mean And Standard Deviation of Predictions by Model. Bold text indicates lowest values for each column.}
  \centering
    \begin{tabular}{|l|l|l|l|l|l|l|l|l|}
    \hline
     & DA Mean & DA STD & 5HT Mean & 5HT STD & pH Mean & pH STD & NE Mean & NE STD \\ \hline
    inception & \textbf{215.696} & \textbf{102.976} & \textbf{242.899} & \textbf{162.223} & 0.054 & 0.039 & \textbf{260.879} & \textbf{121.918} \\ \hline
    mlp\_wide & 276.755 & 208.394 & 272.596 & 246.500 & \textbf{0.049} & 0.046 & 336.086 & 259.398 \\ \hline
    mlp\_big & 278.978 & 205.778 & 277.395 & 250.241 & 0.052 & 0.059 & 321.480 & 229.730 \\ \hline
    eeg\_transformer & 319.458 & 121.367 & 344.235 & 218.203 & 0.073 & 0.054 & 384.339 & 169.795 \\ \hline
    ssvep\_former & 374.564 & 103.709 & 376.539 & 172.568 & 0.061 & \textbf{0.034} & 466.972 & 127.977 \\ \hline
    fcn\_wide & 521.156 & 116.325 & 448.591 & 274.379 & 0.084 & 0.039 & 584.825 & 142.119 \\ \hline
    fcn\_ref & 537.572 & 137.574 & 448.886 & 259.622 & 0.084 & 0.037 & 581.373 & 148.009 \\ \hline
    \end{tabular}
  \label{tab:rmse_table}
\end{table}

\begin{figure}[!htb]
  \centering
  \includegraphics[width=0.3333\textwidth]{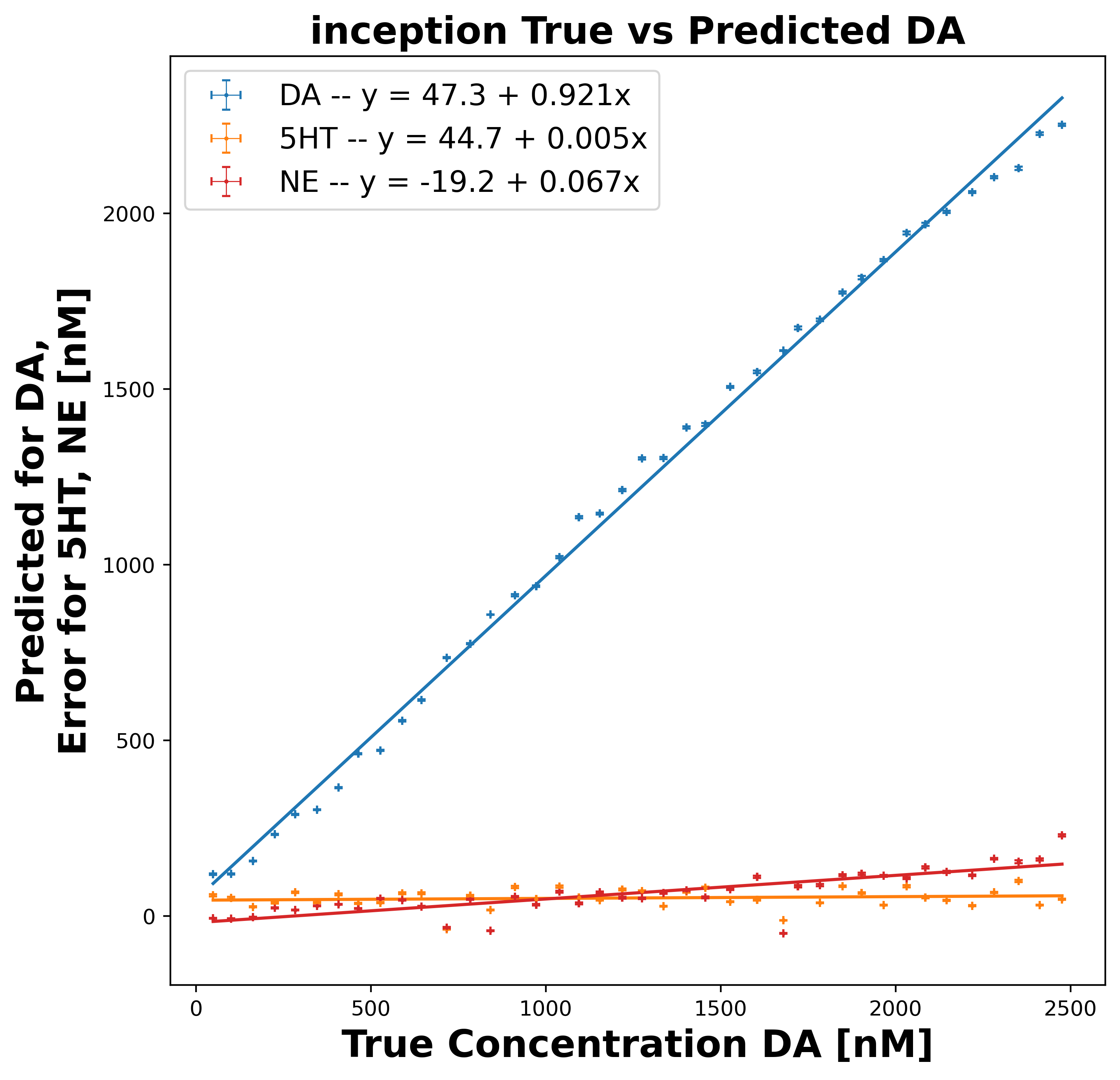}\includegraphics[width=0.3333\textwidth]{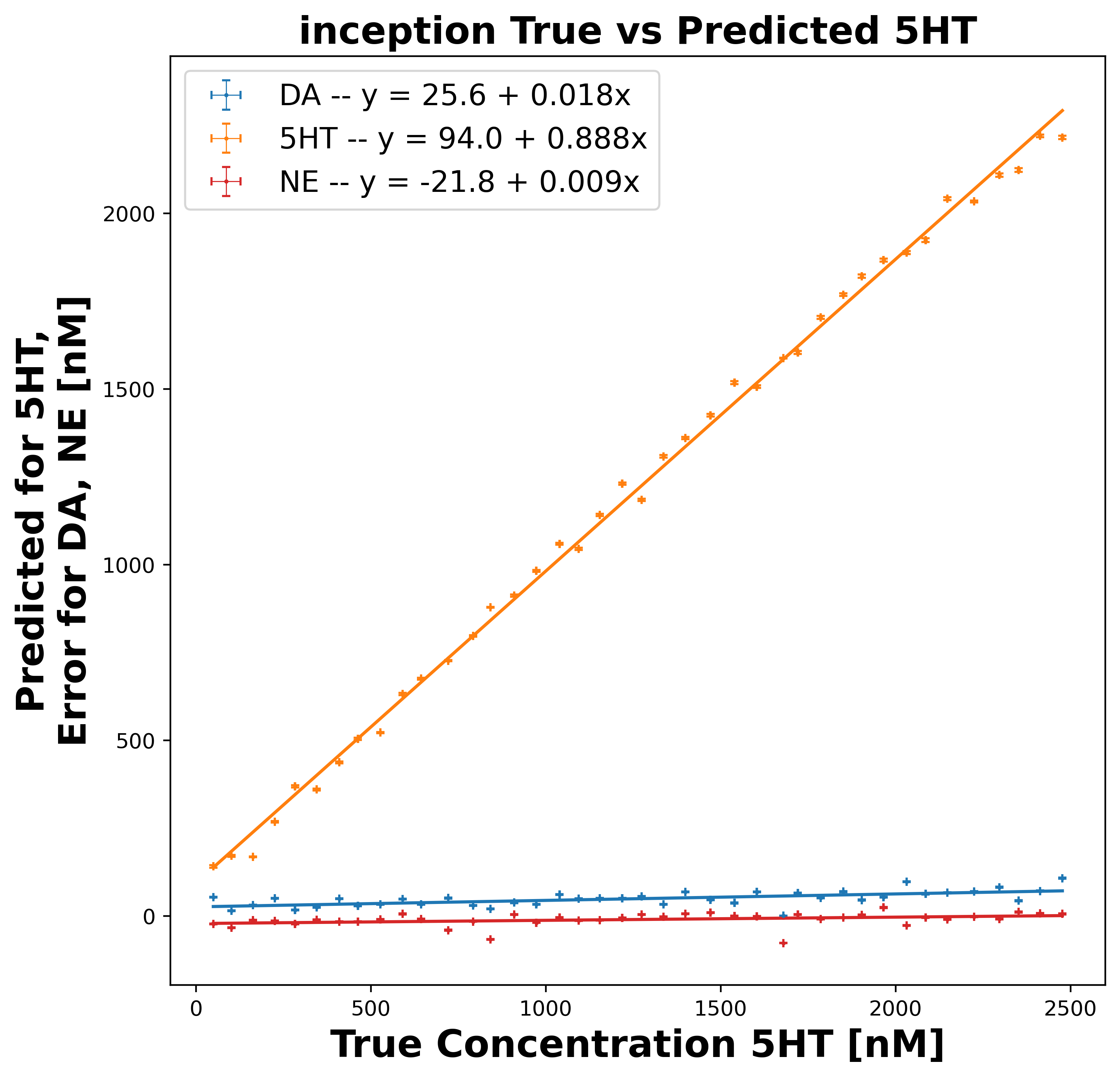}\includegraphics[width=0.3333\textwidth]{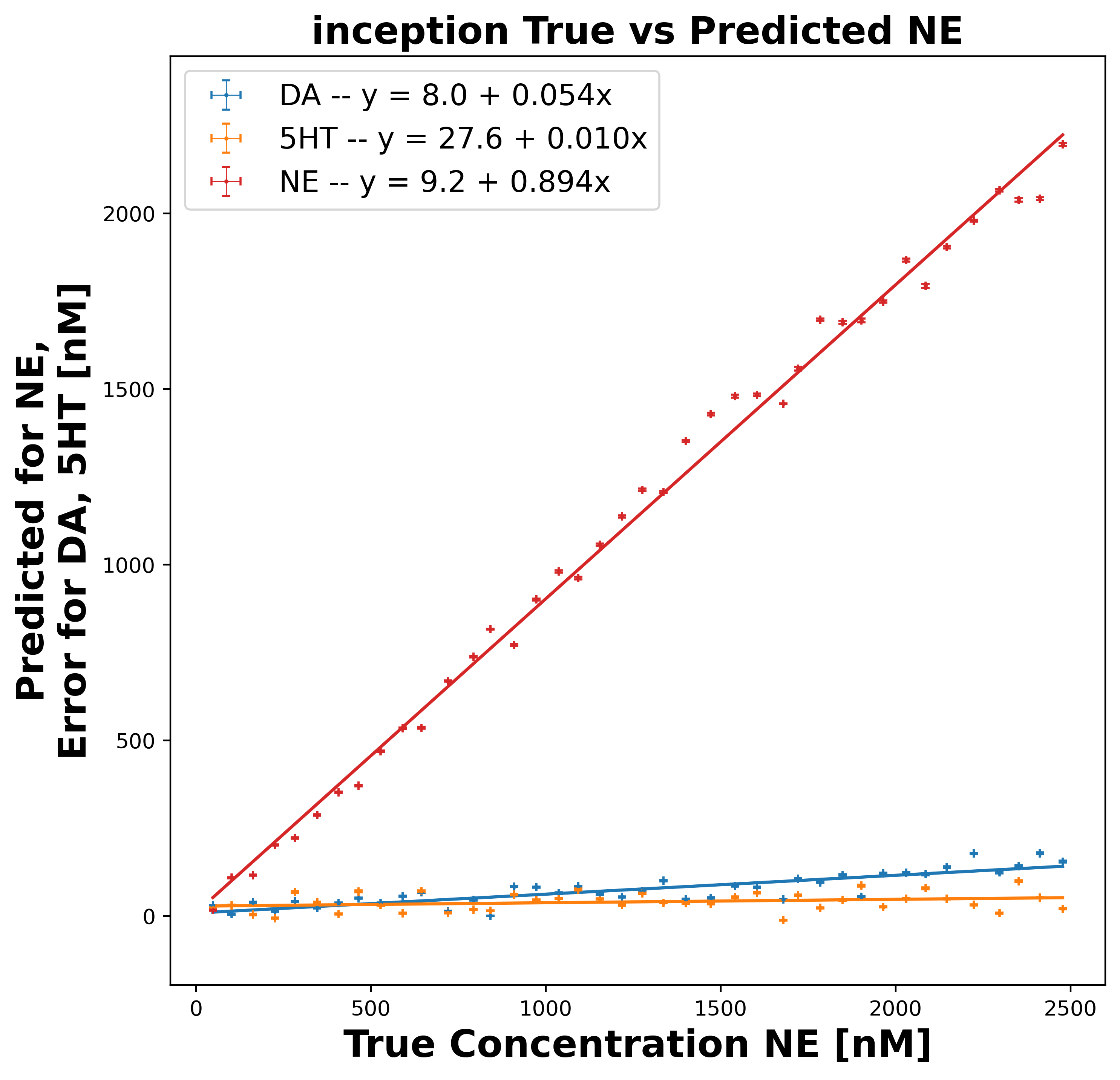}
  \includegraphics[width=0.3333\textwidth]{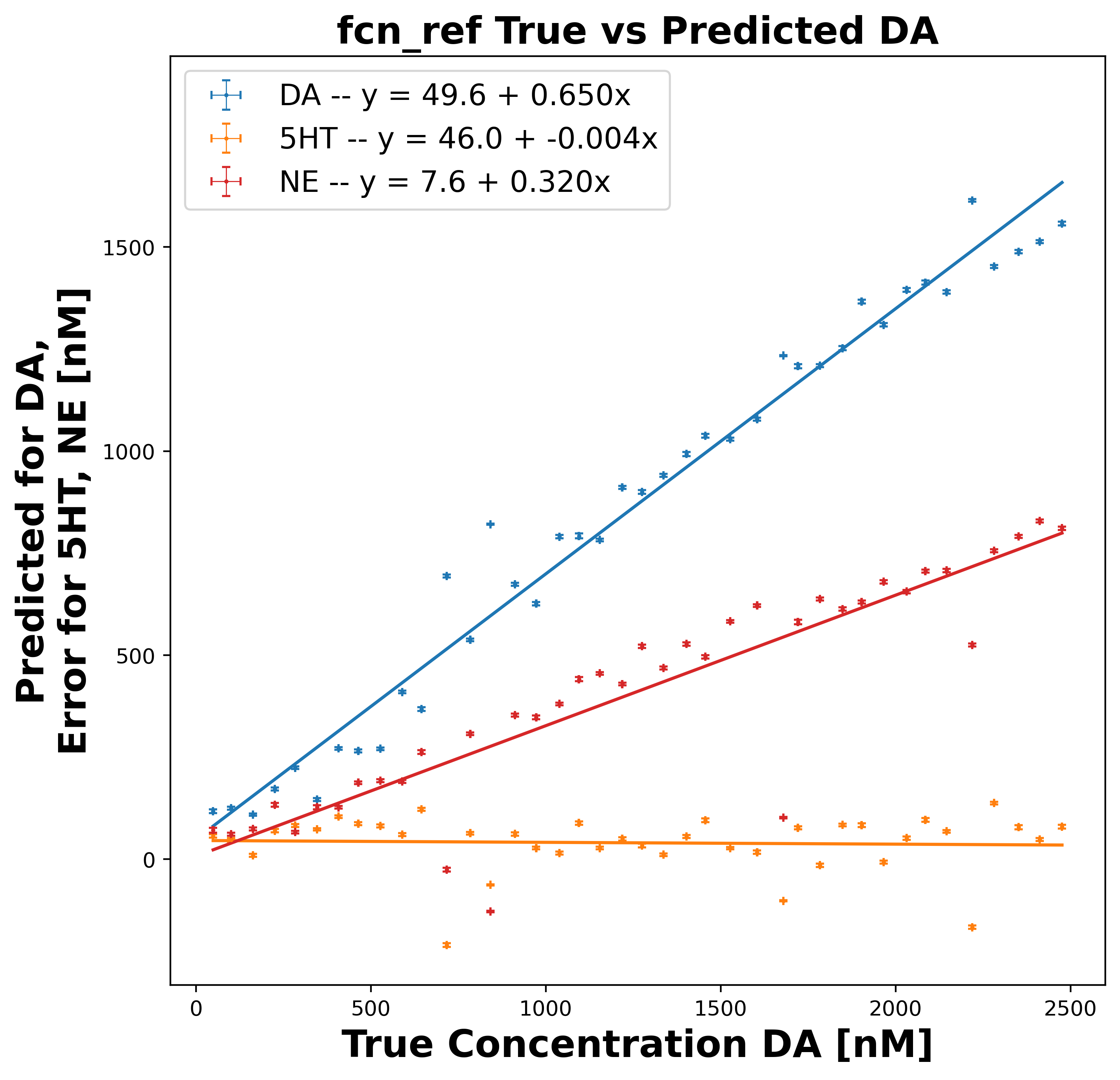}\includegraphics[width=0.3333\textwidth]{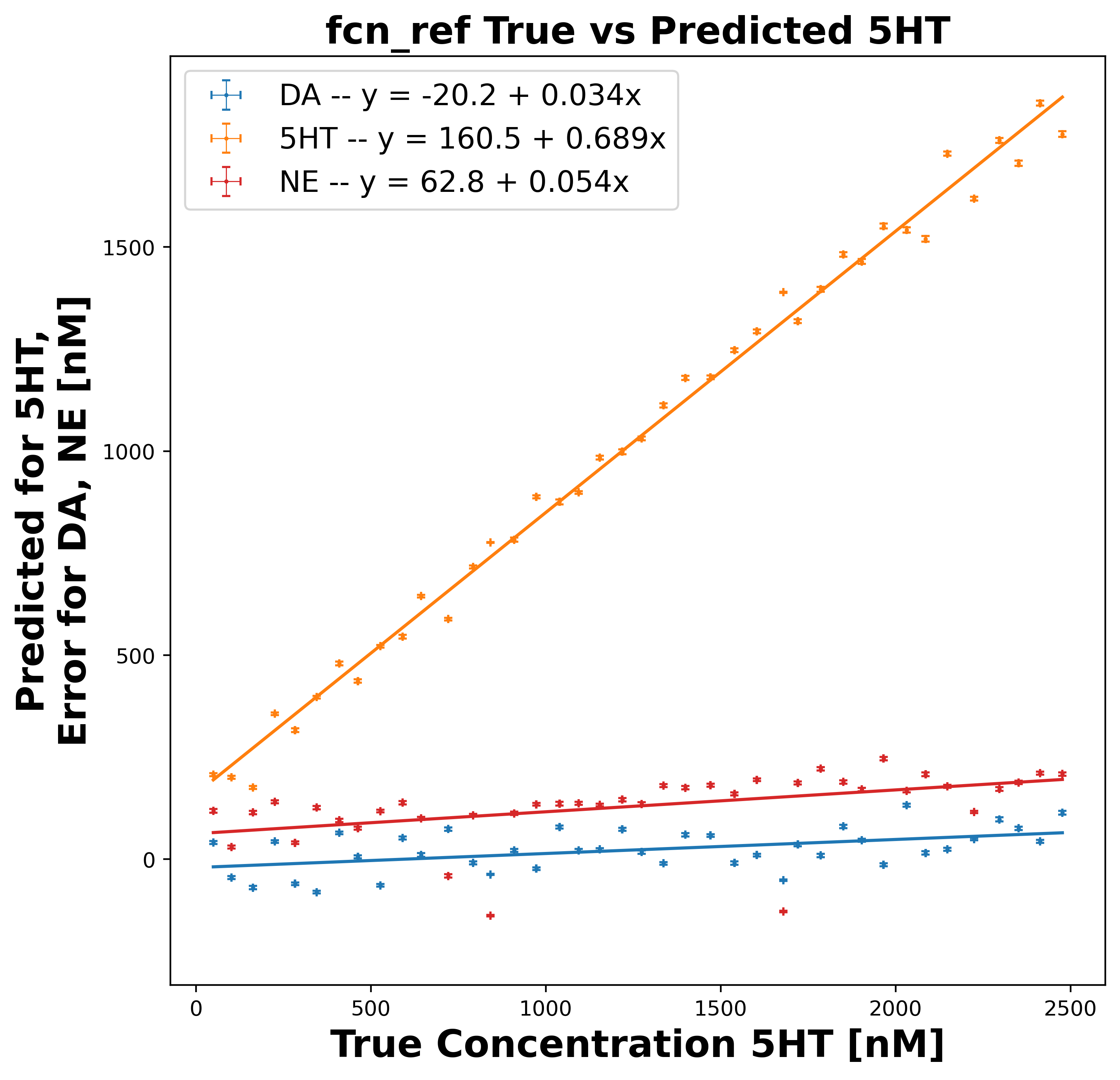}\includegraphics[width=0.3333\textwidth]{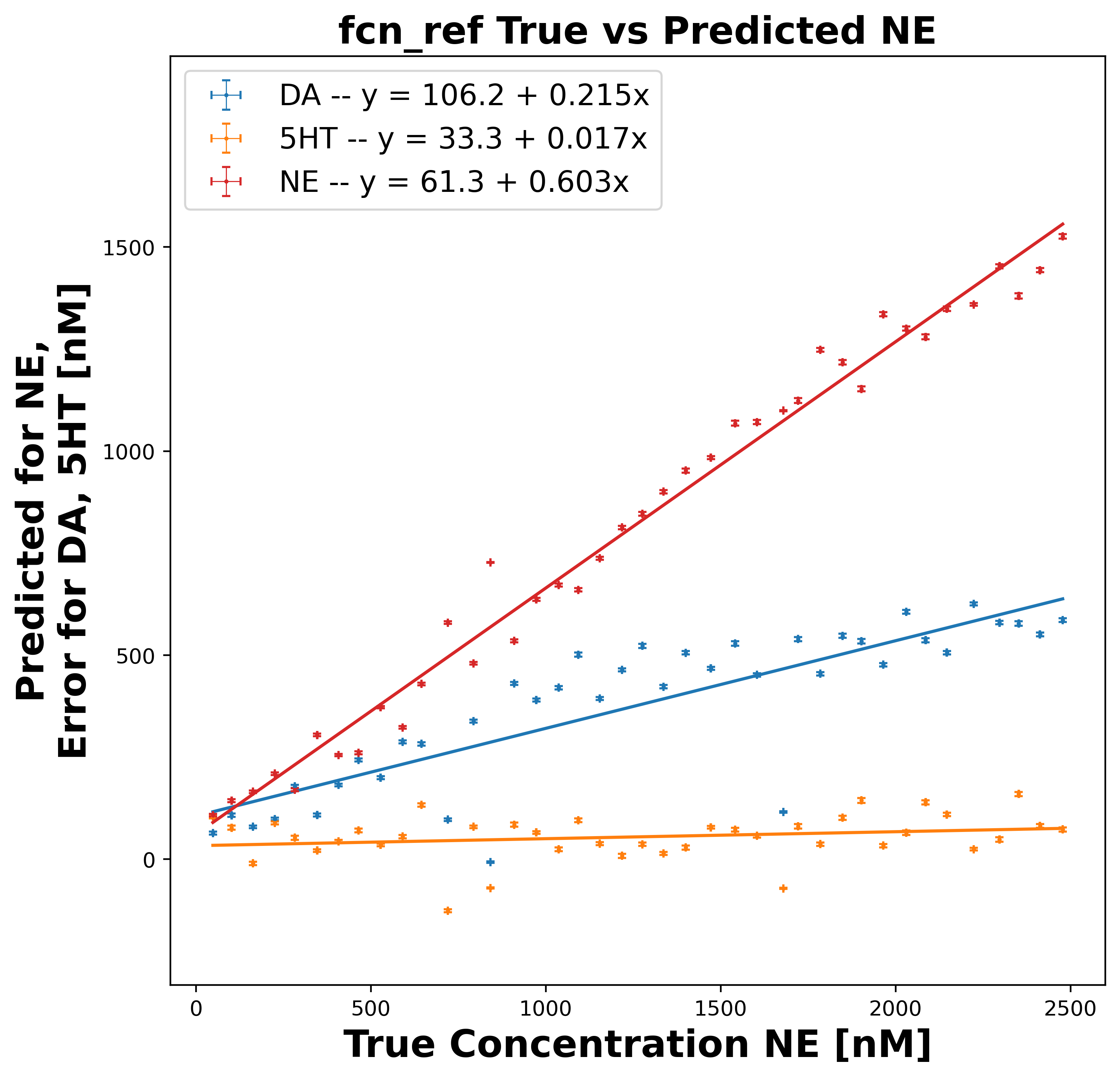}
  \caption{Cross Prediction between analytes for Inception (Top Row) and FCN\_ref models (Bottom Row). For each analyte there is a subplot that shows the relationship between the true and predicted values of that analyte, and error in prediction for the other analytes.}
  \label{fig:cross_true_pred}
\end{figure}

As prior work has had difficulty distinguishing between DA and NE, we examined if each architecture exhibits any “confusion” between these analytes. Figure \ref{fig:cross_true_pred} demonstrates the InceptionTime model exhibiting low confusion. This can be seen as the slope corresponding to the error in the NE as a function of the true NE concentration is much lower than corresponding slope of the DA prediction. The FCN\_wide model exhibits much more “confusion” with the equivalent slopes much closer in magnitude. A similar analysis was performed for all architectures and their corresponding slopes and R-squared values are listed in tables \ref{tab:cross_slope_table} and \ref{tab:cross_r2_table} respectively. The values in the tables come from running the regressions on each test probe and then averaging over the probes whereas the plots combined the probes before running a regression.

\begin{table}[!htb]
 \caption{$R^2$ Values of Cross Predictions}
  \centering
    \begin{tabular}{|l|lll|lll|lll|}
    \hline
    True Analyte & DA &  &  & 5HT &  &  & NE &  &  \\ \hline
    Pred Analyte & \multicolumn{1}{l|}{DA} & \multicolumn{1}{l|}{5HT} & NE & \multicolumn{1}{l|}{DA} & \multicolumn{1}{l|}{5HT} & NE & \multicolumn{1}{l|}{DA} & \multicolumn{1}{l|}{5HT} & NE \\ \hline
    inception & \multicolumn{1}{l|}{\textbf{0.921}} & \multicolumn{1}{l|}{\textbf{0.018}} & \textbf{0.023} & \multicolumn{1}{l|}{0.018} & \multicolumn{1}{l|}{\textbf{0.892}} & 0.020 & \multicolumn{1}{l|}{0.015} & \multicolumn{1}{l|}{0.020} & \textbf{0.882} \\ \hline
    mlp\_wide & \multicolumn{1}{l|}{0.858} & \multicolumn{1}{l|}{\textbf{0.018}} & 0.028 & \multicolumn{1}{l|}{0.020} & \multicolumn{1}{l|}{0.860} & 0.020 & \multicolumn{1}{l|}{0.017} & \multicolumn{1}{l|}{0.023} & 0.798 \\ \hline
    mlp\_big & \multicolumn{1}{l|}{0.846} & \multicolumn{1}{l|}{0.019} & 0.027 & \multicolumn{1}{l|}{0.018} & \multicolumn{1}{l|}{0.859} & 0.019 & \multicolumn{1}{l|}{\textbf{0.013}} & \multicolumn{1}{l|}{0.018} & 0.798 \\ \hline
    eeg\_transformer & \multicolumn{1}{l|}{0.800} & \multicolumn{1}{l|}{0.019} & 0.031 & \multicolumn{1}{l|}{\textbf{0.014}} & \multicolumn{1}{l|}{0.780} & \textbf{0.016} & \multicolumn{1}{l|}{0.019} & \multicolumn{1}{l|}{\textbf{0.017}} & 0.723 \\ \hline
    ssvep\_former & \multicolumn{1}{l|}{0.760} & \multicolumn{1}{l|}{0.034} & 0.094 & \multicolumn{1}{l|}{0.020} & \multicolumn{1}{l|}{0.764} & 0.036 & \multicolumn{1}{l|}{0.045} & \multicolumn{1}{l|}{0.044} & 0.631 \\ \hline
    fcn\_ref & \multicolumn{1}{l|}{0.565} & \multicolumn{1}{l|}{0.027} & 0.118 & \multicolumn{1}{l|}{0.034} & \multicolumn{1}{l|}{0.702} & 0.034 & \multicolumn{1}{l|}{0.097} & \multicolumn{1}{l|}{0.029} & 0.460 \\ \hline
    fcn\_wide & \multicolumn{1}{l|}{0.579} & \multicolumn{1}{l|}{0.023} & 0.116 & \multicolumn{1}{l|}{0.034} & \multicolumn{1}{l|}{0.710} & 0.034 & \multicolumn{1}{l|}{0.091} & \multicolumn{1}{l|}{0.028} & 0.465 \\ \hline
    \end{tabular}
  \label{tab:cross_r2_table}
\end{table}

\begin{table}[!htb]
 \caption{Slope Values of Cross Predictions}
  \centering
    \begin{tabular}{|l|lll|lll|lll|}
    \hline
    True Analyte & DA &  &  & 5HT &  &  & NE &  &  \\ \hline
    Pred Analyte & \multicolumn{1}{l|}{DA} & \multicolumn{1}{l|}{5HT} & NE & \multicolumn{1}{l|}{DA} & \multicolumn{1}{l|}{5HT} & NE & \multicolumn{1}{l|}{DA} & \multicolumn{1}{l|}{5HT} & NE \\ \hline
    inception & \multicolumn{1}{l|}{\textbf{0.908}} & \multicolumn{1}{l|}{-0.103} & -0.054 & \multicolumn{1}{l|}{-0.095} & \multicolumn{1}{l|}{\textbf{0.892}} & -0.110 & \multicolumn{1}{l|}{-0.058} & \multicolumn{1}{l|}{-0.098} & \textbf{0.869} \\ \hline
    mlp\_wide & \multicolumn{1}{l|}{0.878} & \multicolumn{1}{l|}{-0.110} & -0.039 & \multicolumn{1}{l|}{-0.121} & \multicolumn{1}{l|}{0.883} & -0.113 & \multicolumn{1}{l|}{-0.061} & \multicolumn{1}{l|}{-0.120} & 0.824 \\ \hline
    mlp\_big & \multicolumn{1}{l|}{0.867} & \multicolumn{1}{l|}{-0.112} & -0.046 & \multicolumn{1}{l|}{-0.107} & \multicolumn{1}{l|}{0.878} & -0.097 & \multicolumn{1}{l|}{-0.054} & \multicolumn{1}{l|}{-0.087} & 0.834 \\ \hline
    eeg\_transformer & \multicolumn{1}{l|}{0.758} & \multicolumn{1}{l|}{\textbf{-0.057}} & \textbf{0.017} & \multicolumn{1}{l|}{\textbf{-0.065}} & \multicolumn{1}{l|}{0.762} & -0.077 & \multicolumn{1}{l|}{\textbf{-0.011}} & \multicolumn{1}{l|}{\textbf{-0.058}} & 0.723 \\ \hline
    ssvep\_former & \multicolumn{1}{l|}{0.621} & \multicolumn{1}{l|}{-0.099} & 0.121 & \multicolumn{1}{l|}{-0.072} & \multicolumn{1}{l|}{0.605} & -0.096 & \multicolumn{1}{l|}{0.113} & \multicolumn{1}{l|}{-0.101} & 0.477 \\ \hline
    fcn\_ref & \multicolumn{1}{l|}{0.573} & \multicolumn{1}{l|}{-0.089} & 0.203 & \multicolumn{1}{l|}{-0.097} & \multicolumn{1}{l|}{0.710} & \textbf{-0.062} & \multicolumn{1}{l|}{0.187} & \multicolumn{1}{l|}{\textbf{-0.058}} & 0.490 \\ \hline
    fcn\_wide & \multicolumn{1}{l|}{0.583} & \multicolumn{1}{l|}{-0.083} & 0.203 & \multicolumn{1}{l|}{-0.095} & \multicolumn{1}{l|}{0.719} & -0.068 & \multicolumn{1}{l|}{0.182} & \multicolumn{1}{l|}{-0.063} & 0.500 \\ \hline
    \end{tabular}
  \label{tab:cross_slope_table}
\end{table}

\subsection{Response To Noise}
\begin{figure}[!htb]
  \centering
  \includegraphics[width=\textwidth]{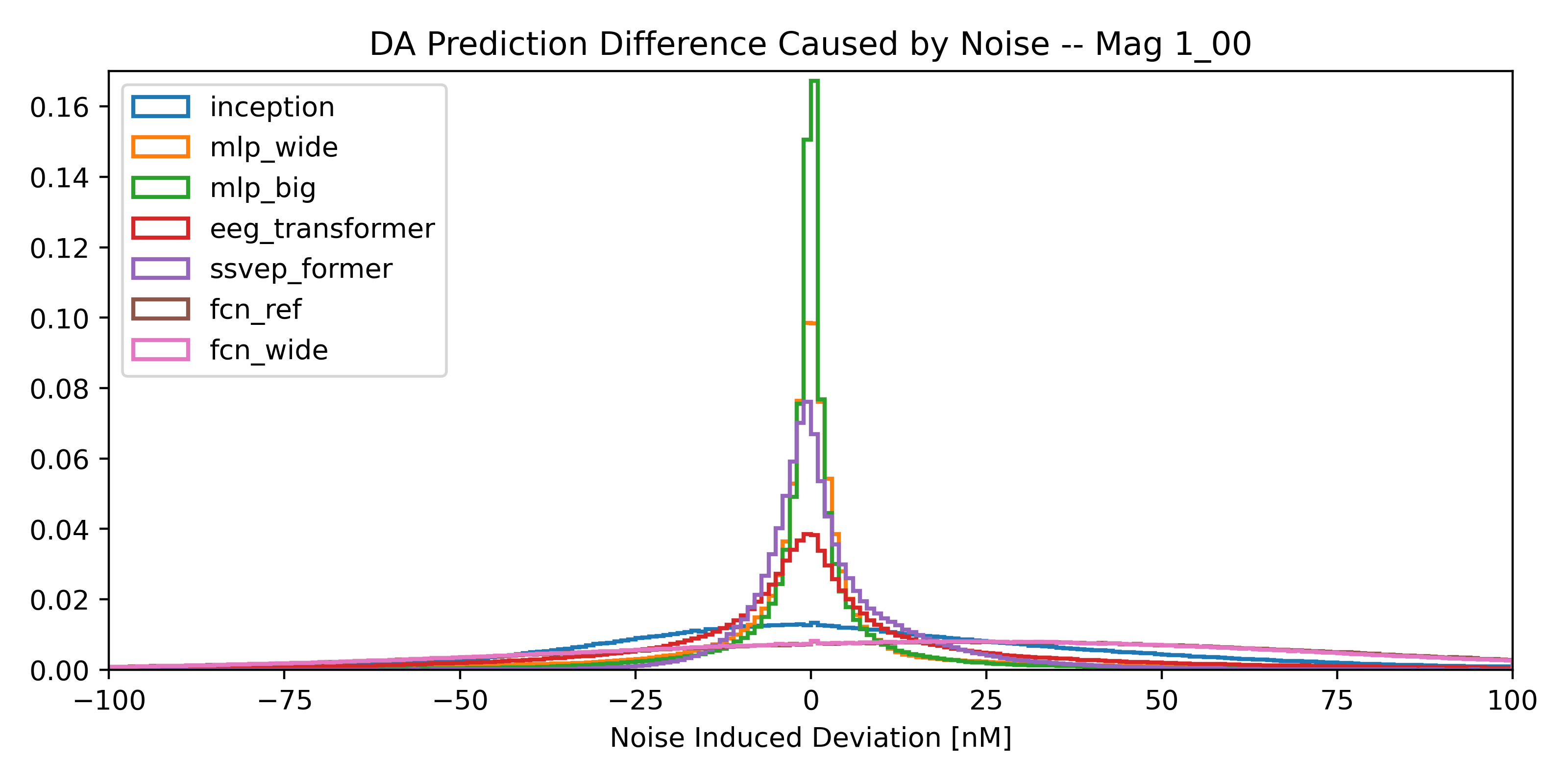}
  \caption{Histogram comparing model's noise response to stir plate noise that has been scaled to 1 nA. The distribution shown is with all of the probes. A log y-scale is used to show separation between models. 1,637,976 sweeps were predicted from each model and are included in the figure.}
  \label{fig:noise_response_hist}
\end{figure}

We consider the difference in the prediction between a base sweep with noise added and the prediction of the base sweep. The distribution of this “noise induced deviation” for the DA predictions over all the probes and their sweeps for the stir plate noise of magnitude 1 nA is shown in Figure \ref{fig:noise_response_hist}. The InceptionTime and both version of the FCN networks are most affected by this noise and seem to have more normal distributions compared to the fatter tailed distributions of the other models. 

\begin{figure}[!htb]
    \centering
    \includegraphics[width=0.5\textwidth]{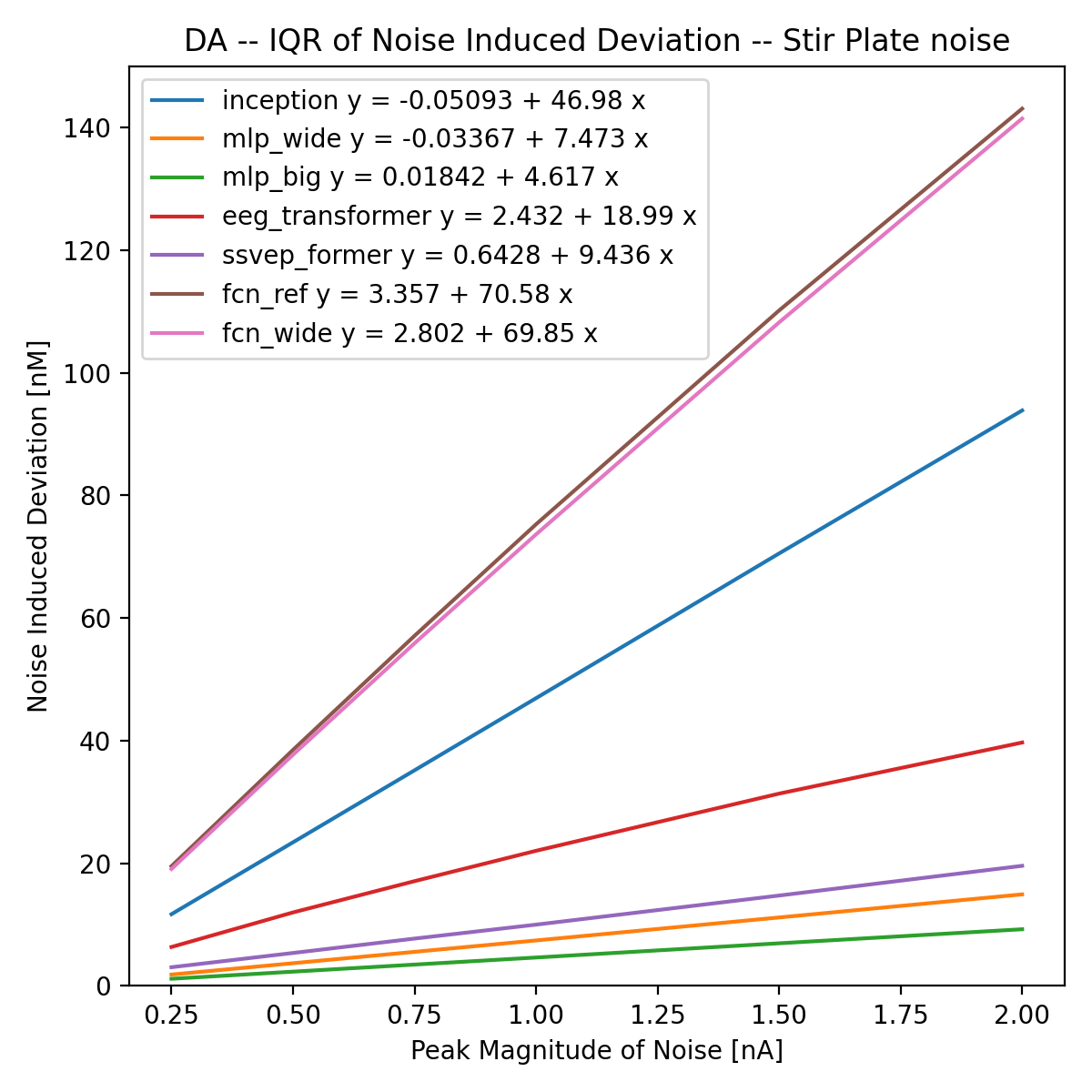}\includegraphics[width=0.5\textwidth]{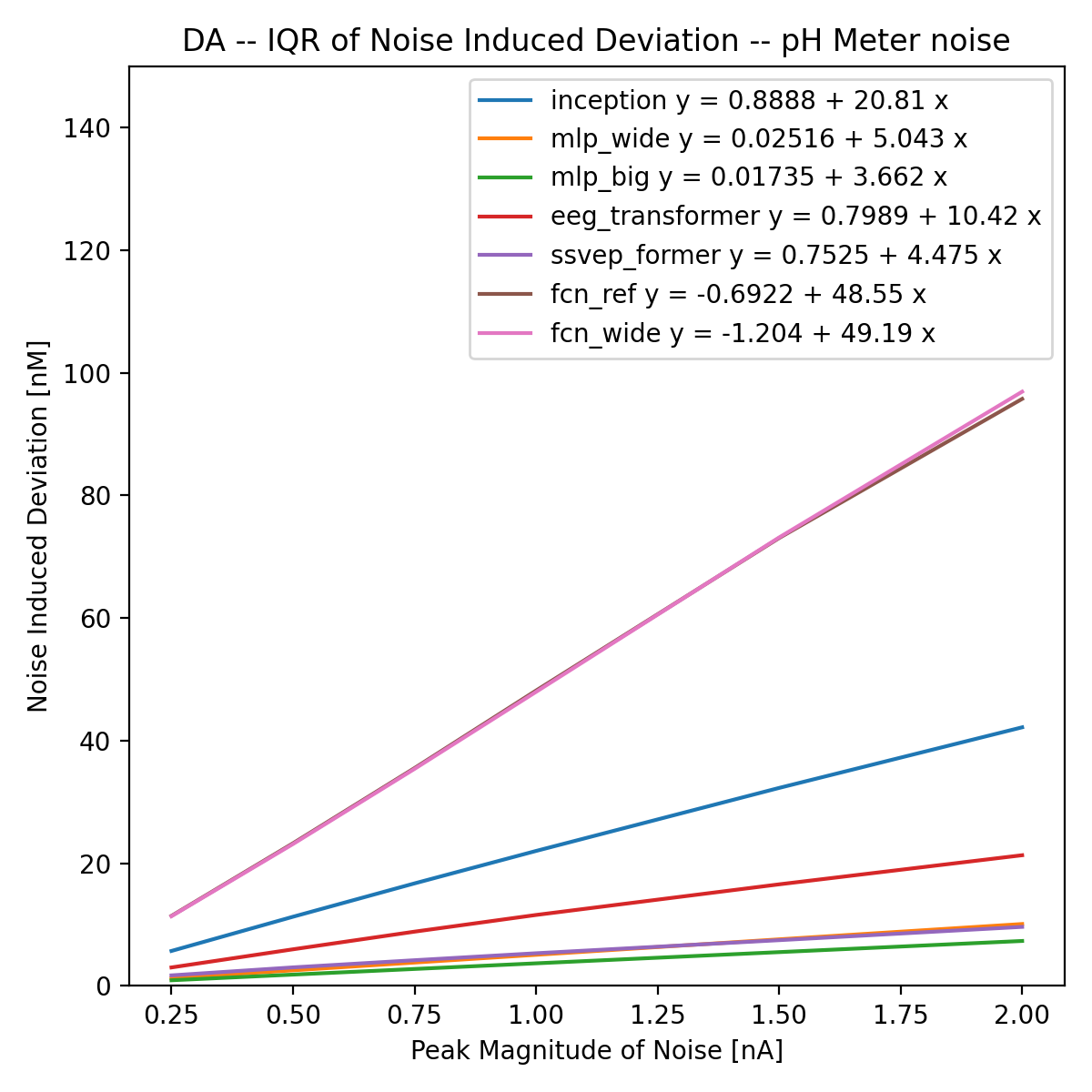}
    \caption{Scaling of interquartile range (IQR) of noise induced deviation with magnitude of noise. Stir Plate noise [Left] and pH Meter noise [Right]}
    \label{fig:noise_mag_scaling}
\end{figure}

For all architectures, the magnitude of the “noise-induced-deviation” scales linearly with the magnitude of the noise added in the range sampled. Figure \ref{fig:noise_mag_scaling} shows this scaling for the DA predictions with both noise sources. Between the two noise sources the order of models in terms of noise effect is unchanged, but the scale differs in both relative and absolute terms. For inception the slope is 2.25 times as great in the Stir Plate noise case, and for the MlP\_wide model it is only 1.5 times as great. The fully convolutional networks are consistently the most affected by either type of noise. A table is provided in the appendix with slopes for all analytes (See tables \ref{tab:Slope_Noise_Scaling_pH_Meter} and \ref{tab:Slope_Noise_Scaling_Sitr_Plate}. 
For a given model the noise response is a function of the base sweep, the specifics of the noise waveform, and the phase of the noise in relation to the sweep. Because of the interplay between those factors, the universe of plausible noise interactions is quite large and the sample we tested may not be representative. 

\subsection{Response To Deviant Probes}

\begin{figure}[!htb]
    \centering
    \includegraphics[width=\textwidth]{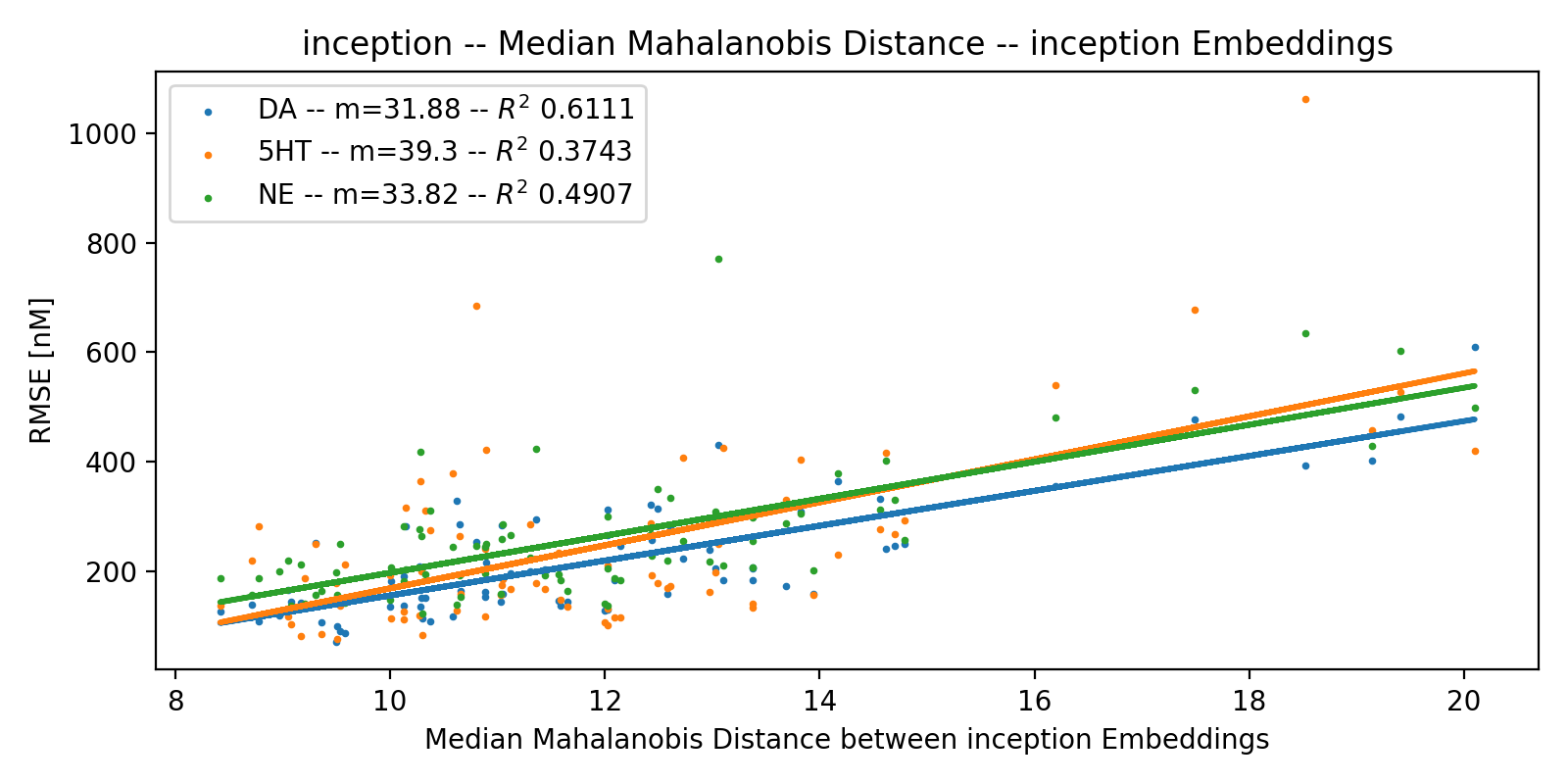}
    \caption{Median Mahalanobis Distance between training distribution and sweep in test probe vs Inception RMSE results. Each dot represents a single analyte on a single probe.}
    \label{fig:m_dist_inception}
\end{figure}

As shown in Figure \ref{fig:boxplots} even models with low median RMSE values fail to accurately predict concentrations for some probes. With these deviant probes, RMSE values can be greater than twice the median. For the inception, MLP\_wide, and eeg-transformer models we looked for metrics correlated with model performance to serve as an indicator of model confidence. We considered the Mahalanobis and Fr\'echet distances between the internal model embedding. To create the internal model embeddings the last layer of each model was removed and the values output from the second to last layer were used. For the Mahalonobis distance, for each test probe the median of the distances from each sweep to the training distribution was used. The Fr\'echet distance compared the distribution of the training set and the distribution of all the sweeps from the test probe. The best distance depends on the model and analyte. Strong correlations exist between median Mahalanobis distance of the inception model embedding and the RMSE performance for DA and and NE. An example is shown in Figure \ref{fig:m_dist_inception} where the DA prediction RMSE had a a $R^2$ value of 0.61. The Fr\'echet distance was better for the MLP\_wide and eeg\_transformer models when using their own embeddings.  As shown in Table \ref{tab:deviant_probe_m_f_dist}, the median Mahalnobis distance on the inception embedding had a stronger correlation with RMSE performance of the model than when using the model's own embedding for both the MLP\_wide and eeg-transformer models. 

\begin{table}[!htb]
    \centering
    \begin{tabular}{|l|l|l|l|l|l|l|l|}
    \hline
    rmse\_model & dist\_model & m\_dist\_DA & m\_dist\_5HT & m\_dist\_NE & f\_dist\_DA & f\_dist\_5HT & f\_dist\_NE \\ \hline
    inception & inception & \textbf{0.611} & 0.374 & \textbf{0.491} & 0.275 & 0.209 & 0.269 \\ \hline
    inception & mlp\_wide & 0.475 & \textbf{0.408} & 0.414 & 0.133 & 0.214 & 0.095 \\ \hline
    inception & eeg\_transformer & 0.008 & 0.012 & 0.035 & 0.212 & 0.432 & 0.273 \\ \hline
    mlp\_wide & inception & 0.197 & 0.314 & 0.149 & 0.161 & 0.249 & 0.091 \\ \hline
    mlp\_wide & mlp\_wide & 0.195 & 0.358 & 0.205 & \textbf{0.635} & \textbf{0.564} & \textbf{0.558} \\ \hline
    mlp\_wide & eeg\_transformer & 0.000 & 0.000 & 0.000 & 0.099 & 0.439 & 0.101 \\ \hline
    eeg\_transformer & inception & 0.538 & 0.302 & 0.466 & 0.158 & 0.160 & 0.194 \\ \hline
    eeg\_transformer & mlp\_wide & 0.234 & 0.310 & 0.350 & 0.012 & 0.114 & 0.094 \\ \hline
    eeg\_transformer & eeg\_transformer & 0.043 & 0.016 & 0.030 & 0.219 & 0.483 & 0.293 \\ \hline
    \end{tabular}
    \caption{$R^2$ values between the root mean squared error (RMSE) performance of the model in rmse\_model column and the distance metric on the embeddings from the model in the dist\_model column. The m\_dist\_DA column is for the $R^2$ corresponding to the median Mahalobois Distance and the RMSE for the analyte DA. The column f\_dist\_NE is for the $R^2$ corresponding to the Fr\'echet distance and the RMSE for the analyte NE. The interior columns follow the same convention.}
    \label{tab:deviant_probe_m_f_dist}
\end{table}

\section{Discussion}

\subsection{Predictive performance}
The InceptionTime architecture performed the best in our test when considering both the absolute RMSE performance and $R^2$ results. Further the performance was more consistent than the second best architecture MLP with roughly half the STD deviation in RMSE across probes. This performance is inline with expectations from prior work and because the other models were constrained to it as a baseline. The MLP performed better than expected and warrants more investigation. Perhaps the MLP architecture is better suited to this task than the time series or image classification tasks used in prior comparisons. Because current evoked by the interaction with the neurotransmitter is fixed in its position in the sweep the utility of convolutions may be less than when these structures are not fixed. Adding another layer to the MLP had little effect on the accuracy suggesting a bottleneck elsewhere in the model or training. The FCN conversely performed worse than suggested by prior work and both sizes of the model had nearly identical results. 

\subsection{Response to Noise}
No prior work had examined the effect of noise in the DNNs applied to a FSCV task, however, it is well documented that small noise additions can cause out-sized effects on DNNs in image classification tasks \cite{API2017}\cite{CommonCorruptions20191}\cite{Generalization2018}. Of the better performing models in terms of accuracy, the InceptionTime model is more responsive to the sample of electrical noise we presented. Again the MLP performs surprising well in this regard. It is unclear why the larger version of the MLP was effected by the noise. Using the numerical first derivative of the data may exacerbate the influence of the noise. With the derivative relatively small noise spikes in relation to the magnitude of the sweep cause large changes in the derivative. Those changes are especially important when the raw current sweep is flat and thus the proportional change of the derivative is more significant than areas of the raw sweep is rapidly changing. 

Given the effect of the noise on all of the top performing architectures, this work suggests effort should be allocated to reducing this noise at the source or in prepossessing if possible. One possible option is to train multiple versions of a given model and ensemble the results. The ensemble approach is recommended for the InceptionTime model but was forgone in this analysis to conserve computational resources.

We note that the noise sources used (Shown in Figure \ref{fig:example_noise}) are not Gaussian, exhibit high auto correlation and seem to have many component spikes. Additionally these noise sources are difficult to remove with naive methods as they vary in time and their component frequencies overlap with the information carrying frequencies of the current sweep. The universe of possible noises to arise in a in vivo environment is very large with the interaction of many possible electronic devices. Thus the noise examination presented here should be thought of as more illustrative of possible noise interaction than representative of a model's response to this universe of possible noises.  

\subsection{Response to Deviant Probes}
The finding that the distance between model embeddings is correlated with performance is expected. This practice of examining distances between model embeddings has been used for similar applications such as the Fr\'echet Inception Distance \cite{FID2017}, which is used to evaluate GANs. In that work the distance  It is unclear why the median Mahanobis distance and Fr\'echet distances are more appropriate for different models and different analytes. The second to last layer of the InceptionTime network may be better suited to serve as embedding because it is of a lower dimensionality than the corresponding layer in the MLP, and has a more evenly distributed embedding than the eeg-transformer. The large differences in RMSE for a given test probe between analytes could be explain by the manner in which data was collected. As most of the true variation in concentration of a given analyte is constrained to a single data set the condition of the probe during that data set or specific noise conditions that existed during its collection may effect the ability of model to predict sweeps from that data set.

\section{Conclusion}

We aimed to leverage the advancements of DNNs developed for other applications for the FSCV task. We investigated the optimal model for specific desirable qualities of this task and particularly the out-of-probe case. Using a large test with much more data than any previous work, we found that the InceptionTime architecture had the best performance in terms of RMSE accuracy. Notably the InceptionTime architecture was also more susceptible to perturbation from electrical noise than other models tested. We found that the a simple MLP had the second best accuracy, and was less effected by noise. All models showed deteriorated performance for deviant probes that were far from the training distribution. Additionally, we have shown that these deviant probes can be detected by finding distance to the training distribution for model embeddings.

\section*{Acknowledgments}
This work was supported by the Wellcome Trust (PRM, 091188/Z/10/Z), NIH-NINDS (PRM, R01-NS092701), NIH-NIMH ( PRM, R01-MH124115; PRM, R01-MH122948), the Swartz Foundation (PRM, 2019-11), and a Virginia Tech Foundation Seale Innovation Award (PRM, FY22). 

\bibliographystyle{unsrt}  
\bibliography{references}  

\newpage

\section*{Appendix}

\begin{table}[!htb]
    \centering
    \begin{tabular}{|l|l|l|l|l|l|l|l|l|}
    \hline
     & DA\_STD & DA\_IQR & 5HT\_STD & 5HT\_IQR & pH\_STD & pH\_IQR & NE\_STD & NE\_IQR \\ \hline
    inception & 17.775 & 20.810 & 22.344 & 22.473 & 0.004 & 0.005 & 22.602 & 24.739 \\ \hline
    mlp\_wide & 10.903 & 5.043 & 13.786 & 7.637 & 0.002 & 0.000 & 15.400 & 6.812 \\ \hline
    mlp\_big & 11.309 & 3.662 & 14.872 & 5.273 & 0.002 & 0.000 & 14.835 & 3.993 \\ \hline
    eeg\_transformer & 20.773 & 10.418 & 22.818 & 7.819 & 0.004 & 0.001 & 28.420 & 10.559 \\ \hline
    ssvep\_former & 5.680 & 4.475 & 7.546 & 6.123 & 0.001 & 0.000 & 6.294 & 5.006 \\ \hline
    fcn\_ref & 38.233 & 48.552 & 29.738 & 36.604 & 0.004 & 0.005 & 36.110 & 47.302 \\ \hline
    fcn\_wide & 38.465 & 49.189 & 29.017 & 36.159 & 0.004 & 0.005 & 34.322 & 44.294 \\ \hline
    \end{tabular}
    \caption{Slope values for the line of best fit for relationship of noise-induced-deviation and magnitude of noise with \textbf{pH meter} Noise. Values are reported for each analyte and both STD and IQR.}
    \label{tab:Slope_Noise_Scaling_pH_Meter}
\end{table}

\begin{table}[!htb]
    \centering
    \begin{tabular}{|l|l|l|l|l|l|l|l|l|}
     \hline
     & DA\_STD & DA\_IQR & 5HT\_STD & 5HT\_IQR & pH\_STD & pH\_IQR & NE\_STD & NE\_IQR \\ \hline
    inception & 39.136 & 46.982 & 30.907 & 33.412 & 0.007 & 0.007 & 43.074 & 49.178 \\ \hline
    mlp\_wide & 23.400 & 7.473 & 12.808 & 7.004 & 0.003 & 0.001 & 27.031 & 9.702 \\ \hline
    mlp\_big & 20.507 & 4.617 & 16.857 & 4.654 & 0.003 & 0.000 & 24.949 & 5.978 \\ \hline
    eeg\_transformer & 33.622 & 18.988 & 36.359 & 12.449 & 0.007 & 0.002 & 45.396 & 23.704 \\ \hline
    ssvep\_former & 11.185 & 9.436 & 10.762 & 8.882 & 0.001 & 0.000 & 11.776 & 10.735 \\ \hline
    fcn\_ref & 53.967 & 70.578 & 44.540 & 55.295 & 0.006 & 0.008 & 57.940 & 76.587 \\ \hline
    fcn\_wide & 53.180 & 69.853 & 43.851 & 54.197 & 0.006 & 0.007 & 54.662 & 72.076 \\ \hline
    \end{tabular}
    \caption{Slope values for the line of best fit for relationship of noise-induced-deviation and magnitude of noise with \textbf{stir plate} Noise. Values are reported for each analyte and both STD and IQR.}
    \label{tab:Slope_Noise_Scaling_Sitr_Plate}
\end{table}

\end{document}